# EEG Brain mapping based on the Duffing oscillator

Mahmut AKILLI[1*], Nazmi YILMAZ[2]
[1*] Istanbul University--Cerrahpasa, Department of Biophysics, Istanbul, Türkiye.
E-mail: akillimahmut@yahoo.com.tr
[2]Koç University, College of Sciences, Department of Physics, Istanbul, Türkiye.

**Abstract**

This study proposes a Duffing oscillator–based measurement framework for detecting frequency components associated with postsynaptic potential–related activity in EEG recordings. The Duffing oscillator is employed as a nonlinear measurement system whose high sensitivity to weak periodic inputs enables robust frequency estimation in the presence of strong noise. Because postsynaptic potentials are embedded in noisy EEG recordings, they are treated as weak signals. Based on the detected weak-signal frequency distributions across EEG channels, topographic maps of brain activity are constructed to visualize spatial variations in neural activation. For this purpose, EEG signals recorded from two musicians and two audience members during a music experiment were analysed. The frequency components of the weak signals were searched within the 4–37 Hz range across all EEG channels. Subsequently, topographic brain maps were generated according to the number of detected weak-signal components. The results indicate that Duffing oscillator–based weak signal detection is an effective tool for EEG brain mapping. Compared with Fourier- and wavelet-based methods, the proposed technique provides a more detailed representation of brain activation. These findings suggest that the proposed approach has potential for investigating various pathological conditions and for enhancing the understanding of cognitive and behavioural functions of the human brain.

## 1. Introduction

Brain mapping is a visual report that analyses the function of the brain by means of non-invasive methods. The non-invasive methods can be divided into two groups according to their principles: electrophysiological and hemodynamic. The electrophysiology group comprises instruments like magnetoencephalography (MEG), transcranial magnetic stimulation (TMS) and electroencephalography (EEG). The hemodynamic group includes tools such as functional magnetic resonance imaging (fMRI), positron emission tomography (PET), single photon emission computed tomography (SPECT) and near infrared spectroscopy (NIRS). [1-5].

Electroencephalography (EEG) is used to record the spontaneous electrical activity of the brain by connecting electrodes on the scalp. In other words, it is the measurement of the electrical potentials of cortical neuronal dendrites adjacent to the brain surface. The process of processing digitized EEG data recorded with multiple electrodes into a computer is called quantitative electroencephalography (qEEG). The quantitative EEG (qEEG) technique focuses on frequency-power spectral analysis of multi-channel EEG data processed using mathematical methods such as Fourier and Wavelet analysis, and EEG data is analysed statistically. Brain mapping is obtained by grading the activity levels of brain regions with colours according to the results of the processed EEG data. For this reason, qEEG techniques are called as “EEG brain mapping”. It aims to obtain information about the cognitive and behavioural functions of the human brain via qEEG techniques. [6-9].

EEG devices do not only record the electrical potentials of the cerebral cortex in the brain. Brain signals are contaminated with noise during recording due to the prevalence of contaminating factors such as action potentials from scalp muscles, electrocardiogram, sweat, voltages caused by eye movements, or poor contact of the electrode with the scalp. However, the qEEG accepts noisy EEG data as if it were real brain signal and processes the entire signal through its analysis and turns it into images, or produces statistical results. Therefore, processing noisy EEG data causes many problems. [10-11].

The Duffing oscillator system is a successful means of detecting weak periodic (or quasi-periodic) signals which have a significantly low signal-to-noise ratio. [12-13]. Therefore, the Duffing oscillator system can be utilized to measure the frequency values of postsynaptic potentials, the primary components underlying EEG signals [14-15]. Due to their embedding in background EEG noise, these postsynaptic potentials are considered weak signals, as their amplitudes are small relative to the noise level. Recently, an automatic method for detecting weak signals in empirical signals has been proposed for the Duffing oscillator [16].

In this study, we propose a brain mapping approach based on the frequency distribution of postsynaptic potential measured in EEG signals using the Duffing oscillator. To do this, we used EEG data obtained from participants in a music experiment [17]. In this music experiment, EEG recordings of musicians and audience members were taken while a section of "Shepherd on the Rock" by Franz Schubert was performed. In our study, we used the EEG signals of two musicians playing the piano and flute and two audience members. These music EEG signals were scanned in the frequency range of 4-37 Hz via the Duffing oscillator. We plotted the spectra of the detected weak signal (or postsynaptic potential) frequencies in all EEG channels for two musicians and two audience members. Then, brain topographical maps of the participants were made based on the numbers of these weak signals detected, as shown in **Figure 1**. This technique was compared with those derived from Fourier- and wavelet-based mapping methods. Can brain mapping based on the Duffing oscillator reveal significant differences in the EEG signals of musicians and audiences? If the answer to this question is positive, it will encourage the use of this technique in research on behavioural and cognitive brain functions.

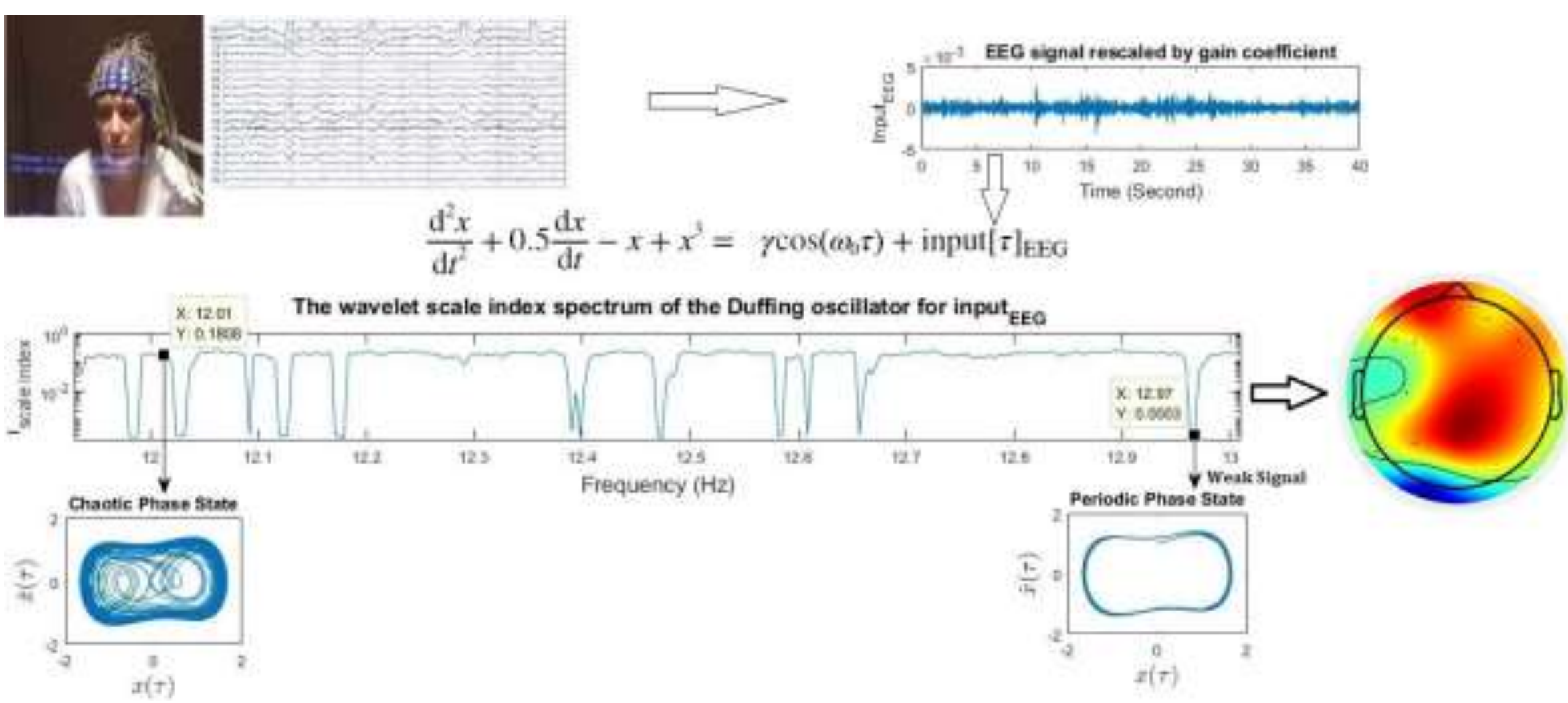

**Figure 1:** Graphic abstract. Detecting weak signals in EEG data and the brain mapping process.

## 2. Material and Method

### 2.1. Electroencephalography (EEG) Data

Raw EEG signals were taken from the music experiment of a previously published study [17], https://doi.org/10.3389/fpsyg.2018.01341 . This musical experiment performed an excerpt from "Shepherd on the Rock" by Franz Schubert in front of a group audience members. EEG signals of musicians and audience members were collected during this live chamber music concert. The EEG sampling frequency is 250 Hz. The EEG device is band-pass filtered between 2 Hz and 40 Hz. Our study utilized EEG signals from two audience members and two musicians. Here, two musicians were playing via piano and flute in the 'strict' mode of the piece. One of these two audience members is male and the other is female. Both of them had low engagement with music. He could both see and hear the musicians, she could only hear, but could not see them. Electrode positions of the International 10-20 system were used for the recording of the EEG (Cz, C3, C4, Fp1, Fp2, Fz, F3, F4, F7, F8, O1, O2, Pz, P3, P4, P7, P8, T7, T8). The ground electrode was positioned on the forehead and the reference electrode behind Cz. 18 electrodes were used on the flutist, 12 on the pianist, 19 on the male audience, and 18 on the female audience. [17].

### 2.2. Duffing Oscillator as Weak Signal Scanner

The Duffing oscillator system is a sensitive means of finding weak periodic or quasi-periodic signals that may be hidden in a noisy signal. This is because the Duffing system is on the edge of chaos and when forced by a weak external signal the system transitions to its periodicity stability. Therefore, the weak signals within highly noisy data can be detected by scanning the whole frequency domain via the Duffing oscillator system. [12-13]. The form of the Duffing equation (1), when an input signal is included,

$$\frac{d^2x}{dt^2} + 0.5\frac{dx}{dt} - x + x^3 = \gamma\cos(t) + \text{input} \tag{1}$$

Frequency transformation is done to detect weak signals of unknown frequency in the input signal. Defining $t = \omega_0\tau$, where $\omega_0$ is the angular frequency. Then, the Duffing equation (1) is modified [12];

$$\begin{aligned} \frac{\mathrm{dx}}{\mathrm{d\tau}} &= \dot{\mathrm{x}}(\tau) = \omega_0 \mathrm{y} \\ \frac{dy}{d\tau} &= \dot{\mathrm{y}}(\tau) = \omega_0[-0.5\mathrm{y} + \mathrm{x} - x^3 + \gamma\cos(\omega_0\tau) + input(\tau)] \end{aligned} \tag{2}$$

Where $\gamma\cos(\omega_0\tau)$ is the reference signal, $\gamma$ is its amplitude. The amplitude value $\gamma$ that allows the Duffing system to transition from critical chaos to periodic stability is called the bifurcation value $\gamma_c$. The input signal is a noisy signal in which weak signals are embedded. Firstly, before the input (or external) signal is added to the Duffing equation, as shown in **Figure 2(a),** the Duffing system is brought into the 'critical chaotic phase state' by setting the reference signal's amplitude $\gamma$ very close to its bifurcation value $\gamma_c$. Secondly, in order to detect weak signals, the input signal is scanned by varying the angular frequency $\omega_0$ in the Duffing equation (2). During the frequency domain scan, when the reference frequency ($\omega_0$) captures the 'weak signal frequency' ($\omega_0$=$\omega$), the Duffing system switches to the periodic phase, as demonstrated in **Figure 2(b)** [12-13].

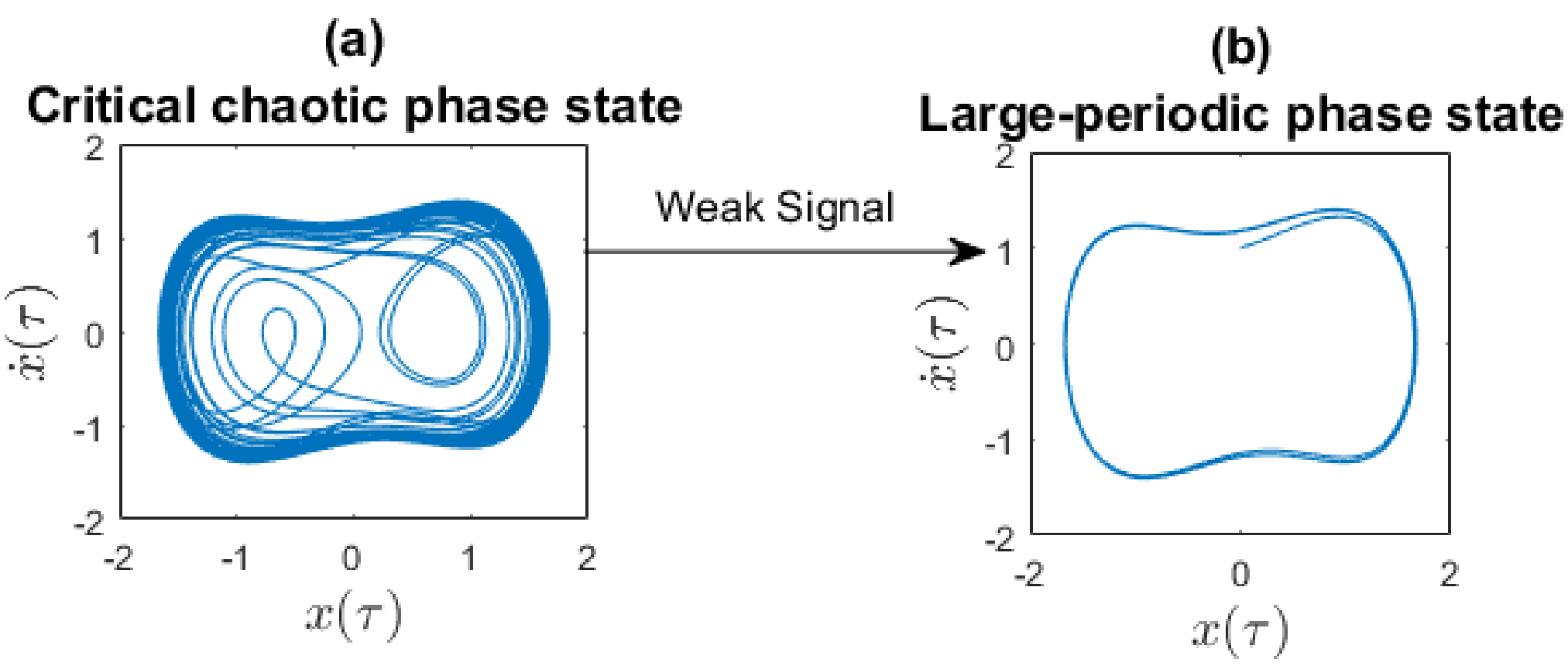

**Figure 2:** Phase states of the Duffing system. (**a**) at the edge of chaos (**b**) at periodic state.

In a recent study [16], the frequencies of the weak signals detected in the input signal were displayed on the Duffing oscillator's scale index spectrum. The scale index rates the aperiodic state of a signal in the range of [0,1] using the wavelet scalogram [18-20]. The scale index value approaches zero for periodic signals. The wavelet scale index values of the time series $x(\tau)$ derived from equation (2) of the Duffing chaotic oscillator are calculated for each angular frequency value $\omega_0$. **Figure 1** demonstrates that the scale index values of the periodically stable phase of the Duffing oscillator correspond to values less than $i_{scale} < 10^{-2}$ in its scale index spectrum [15-16, 18,21].

**Example:**

In practice, how can the Duffing oscillator be used to find weak periodic signals (of a periodic or quasi-periodic nature) in EEG data?

**Figure 3(a)** shows the EEG data from the first 40 seconds of the Fp1 channel recording of a musician playing "Shepherd on the Rock" by Franz Schubert in 'strict' mode through the **piano** [17].

In order to use the EEG data as input signal in the Duffing equation (2) [16];

1) The EEG data is rescaled by multiplying it by an appropriate gain coefficient to fit the reference signal scale, $input_{EEG} = c * EEG\ signal$. Because the Duffing system crashes when the input signal amplitude is substantially larger than the reference signal and it is not excited when the input signal amplitude is very small. [16]. **Figure 3(b**) shows that the music EEG data was multiplied by an appropriate gain coefficient of 0.000025.
2) The sampling rates (or frequencies) of the Duffing system and the input signal should be set equal to each other, $f_s^D = f_s^{input}$ [16]. In the music experiment, EEG signals were collected at a sampling frequency of $f_s^{EEG}$=250 Hz [17]. Their sampling rates were set to 1000 Hz, $f_s^D = f_s^{input} = 4f_s^{EEG} = 4 \times 250 = 1000\ Hz$. Then, the detected weak signal frequencies were calculated by dividing by 4, $\frac{f_s^{input}}{f_s^{EEG}}$. The Duffing equation (2) was resolved through numerical means using the Runge Kutta 4th order method [22] with a step size of $h = \frac{100}{f_s^D} = \frac{100}{1000} = 0.1$

Before adding the input signal shown in **Figure 3(b)** to the Duffing equation (2), the reference signal's amplitude was set to be very close to its bifurcation value, $\gamma = 0.82556$. The Duffing system attains periodic-stable state when the reference amplitude value $\gamma_c$ equals 0.82557. Next, the EEG data was scanned in the angular frequency range of $3 \leq \omega_0 \leq 3.519$ with the step of 0.001 via the Duffing oscillator. The relationship between angular frequency, denoted by $\omega_0$ and frequency, denoted by $f$, for the Duffing oscillator is given by the formula $f = \frac{100\omega_0}{2\pi} = 4f_{EEG}$. Thus, EEG frequency values $f_{EEG}$ fall within the span of 11.93 Hz to 14 Hz.
The scale index values of the time series $x(\tau)$ derived from equation (2) for this angular frequency range ($\omega_0$) were calculated by employing the Haar wavelet with the scales ranging from $s_0 = 1$ to $s_1 = 512$. In **Figure 3(c)** the weak signals with a periodic or quasi-periodic nature are indicated by the scale index values below $i_{scale} < 10^{-2}$.

The weak signal amplitudes detected by the Duffing oscillator are $a \geq \gamma_c - \gamma = 0.82557 - 0.82556 \geq 0.00001$. The real values of the weak periodic signal amplitude ($a'$) detected in the signal are calculated by dividing them by the gain coefficient, $a' \geq \frac{a}{0.000025} = \frac{0.00001}{0.000025} \geq 0.4\ \mu V$. Accordingly, weak signals with an amplitude values of $a' \geq 0.4\ \mu V$ can be detected in the EEG signal. Weak signal (W) to EEG signal (S) ratio (SWR):

$a' = 0.4\ \mu\text{V},\ (\text{Fp1})_{\text{rms}} = 10.994\ \mu\text{V};$

$$\mathrm{SWR} = 10\log_{10}\left(0.5\frac{(a')^2}{((\mathrm{Fp1})_{\mathrm{rms}})^2}\right) = 10\log_{10}\left(0.5\frac{(0.4)^2}{(10.994)^2}\right) = -31.79\mathrm{db}$$

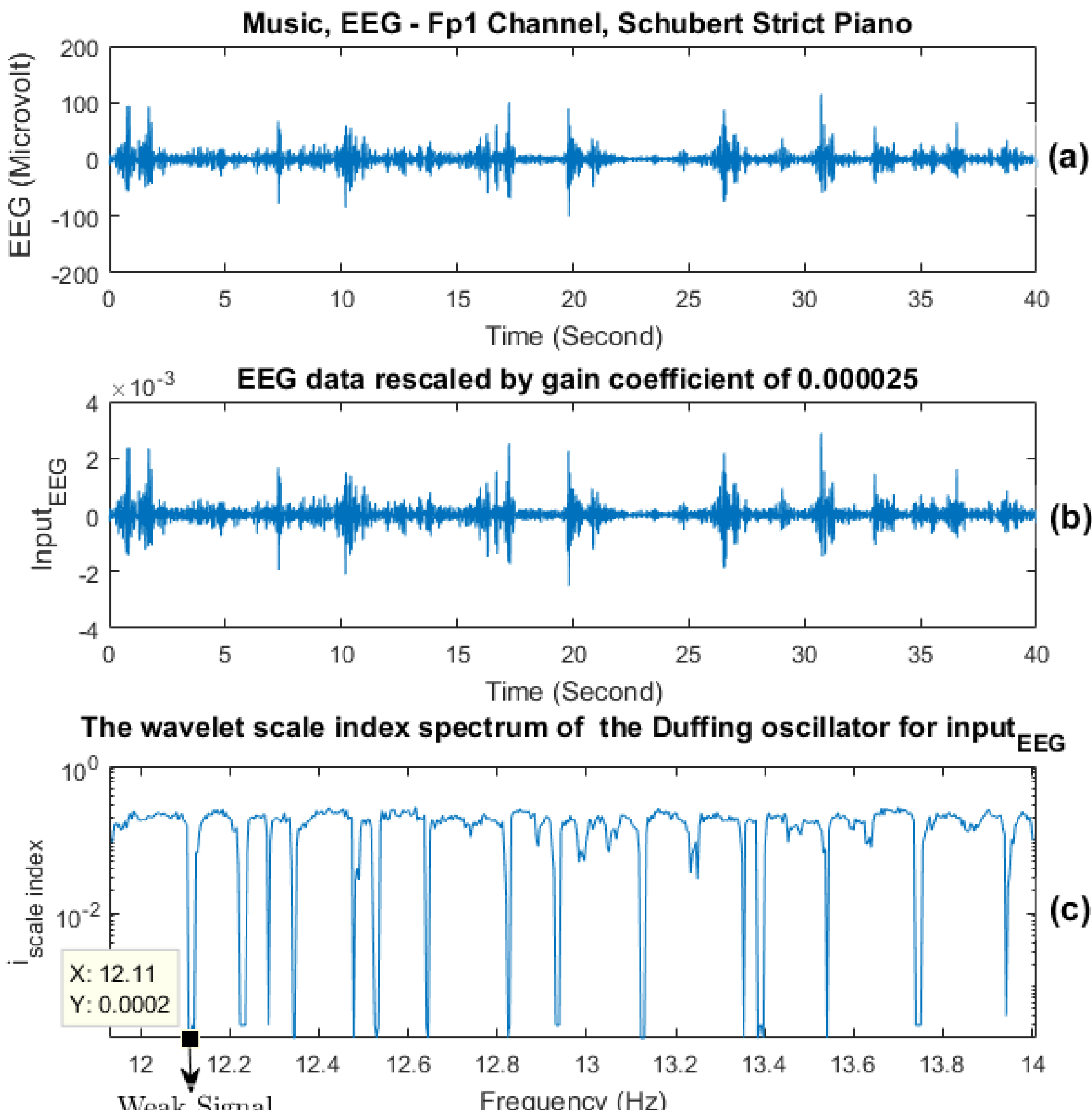

**Figure 3**: (**a**) The first 40 seconds of the EEG - Fp1 channel recording of a musician performing Franz Schubert's "Shepherd on the Rock" in 'strict' mode via the **piano** [17]. (**b**) EEG data were rescaled by 0.000025. (**c**) Spectrum of wavelet scale index parameters plotted against Duffing oscillator frequency: Frequencies corresponding to scale index values falling below $i_{scale} < 10^{-2}$ denote weak periodic signals within the EEG signal.

## 3. Results

### 3.1. The Spectrum of the Weak Signal Detected in EEG signals

The EEG signals of two musicians (a pianist and a flutist) and two audience members (a male and a female) [17] were scanned by means of the Duffing oscillator. The aim was to find weak signals in these EEG signals. The first 40 seconds of the EEG data were used for this study. Therefore, the duration of detected weak signals is 40 seconds minimum. Making music is a high-frequency activity for the brain. Therefore, we looked for weak signals at frequencies between 4 Hz and 37 Hz.

To use Duffing oscillator as signal scanner; firstly, the Duffing oscillator's sampling frequency $f_s^D$ was set to 1000 Hz. Therefore, the input signal's sampling frequency $f_s^{input}$ was taken as $f_s^D = f_s^{input} = 4 * f_s^{EEG} = 4 * 250 = 1000$ Hz [16]. Here, the EEG signals' sampling frequency $f_s^{EEG}$ was 250 Hz [17]. Secondly, EEG signals were rescaled with a gain of 0.000025 before being inserted as the input signal in the Duffing equation (2).

Before adding EEG input signals to the Duffing equation (2), the reference amplitude $\gamma$ is chosen very close to its bifurcation value $\gamma_c$, which is the reference amplitude that allows the Duffing oscillator to pass from critical chaos to periodic. In the Duffing oscillator system, raising the angular frequency $\omega_0$ also raises the bifurcation value $\gamma_c$. Hence, the amplitude values γ were selected from below intervals:

For $1.0 \leq \omega_0 \leq 1.069$, the amplitude value set to $\gamma = 0.82552$ ($\gamma_c = 0.82553$).

For $1.070 \leq \omega_0 \leq 1.244$, the amplitude value set to $\gamma = 0.82553$ ($\gamma_c = 0.82554$).

For $1.245 \leq \omega_0 \leq 1.520$, the amplitude value set to $\gamma = 0.82554$ ($\gamma_c = 0.82555$).

For $1.521 \leq \omega_0 \leq 2.100$, the amplitude value set to $\gamma = 0.82555$ ($\gamma_c = 0.82556$).

For $2.101 \leq \omega_0 \leq 4.550$, the amplitude value set to $\gamma = 0.82556$ ($\gamma_c = 0.82557$).

For $4.551 \leq \omega_0 \leq 9.30$, the amplitude value set to $\gamma = 0.82557$ ($\gamma_c = 0.82558$).

To detect the weak signals, the EEG input signals were scanned in the angular frequency range of $1.0 \leq \omega_0 \leq 9.3$ with the step of 0.001. As the input signal sampling frequency is 4 times the EEG signal sampling frequency, the EEG frequency values correspond to the range of 3.98 Hz to 37 Hz.

The weak signal amplitudes to be identified by the Duffing oscillator are $a \geq \gamma_c - \gamma = 0.00001$**.** The real amplitude values of the weak signals $a'$ was obtained by dividing it by the scaling factor, $a' \geq \frac{a}{0.000025} = \frac{0.00001}{0.000025} \geq 0.4\ \mu V$. Accordingly, weak signals with an amplitude values of $a' \geq 0.4\ \mu V$ were detected. The Duffing chaotic oscillator's scale index values were computed by the Haar wavelet function scaled from $s_0 = 1$ and $s_1 = 512$.
**Figures 4 and 5** show the spectrum of the weak signal that was detected in the EEG signals of two musicians (a piano player and a flute player). **Figures 6, 7 and 8** show the spectrum of weak signals detected in the EEG signals of two members of the audience (one male and one female). In **Figures 9** and **10**, the weak signal spectrum of the pianist's and audience-male's Fp1 EEG channels are compared using Fourier- and wavelet-based approaches. **Figures 9(d)** and **10(d)** present the power spectrum of the Fp1 EEG signal estimated using the FFT–Welch method, while **Figures 9(e)** and **10(e)** depict the time-averaged wavelet power spectrum derived from the wavelet scalogram [23].

The Kernel Density Estimation (KDE) [24] can be used to facilitate a comparison of the spectra of the weak signals detected in the EEG channels. **Figure 11** shows a comparison of the KDE of weak signals detected in channels O1, P3, F3 and T8 for musicians and audiences.

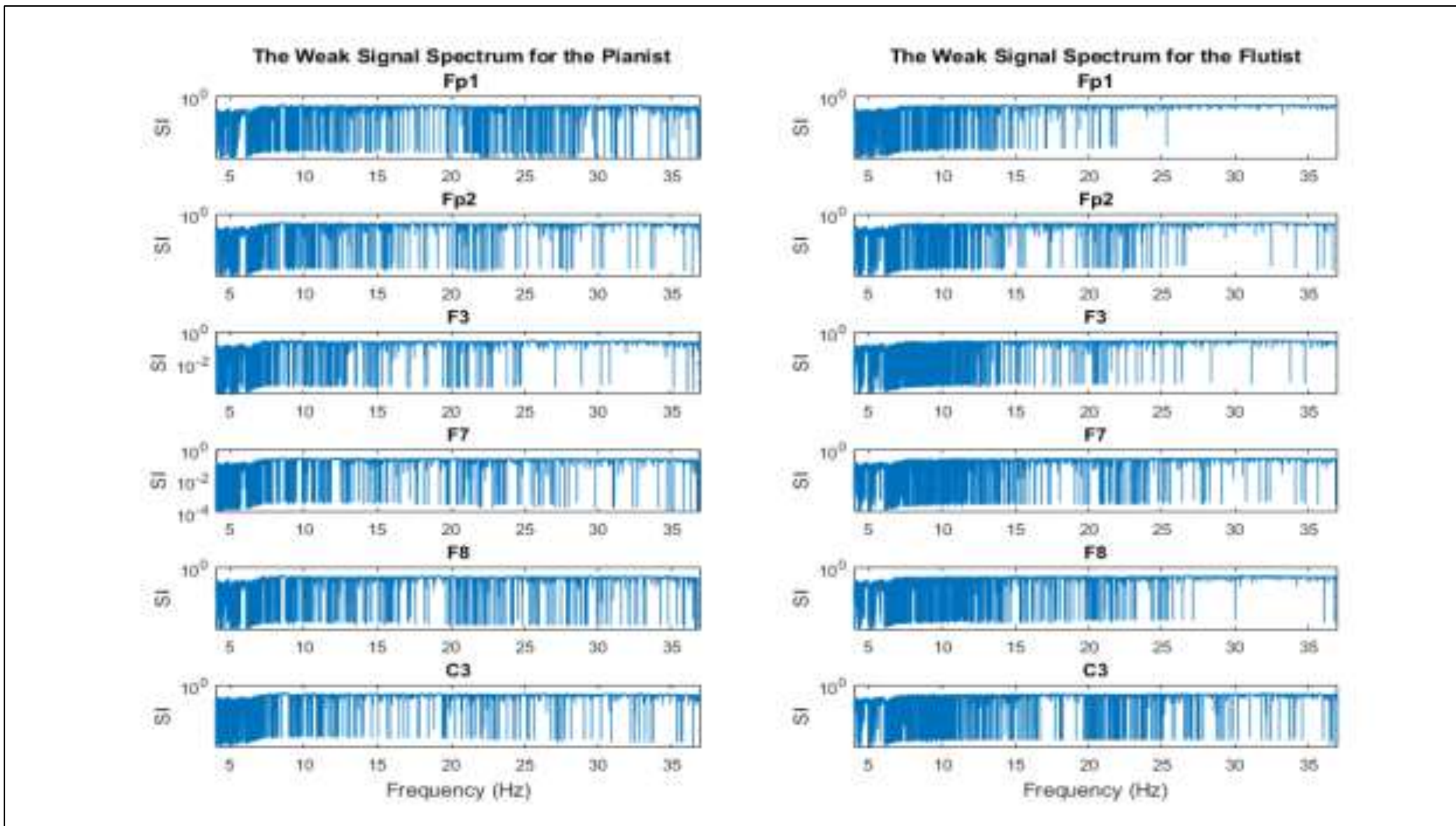

**Figure 4**: Spectrum of wavelet scale index parameters plotted against Duffing oscillator frequency: The EEG signals of two musicians performing Franz Schubert's "Shepherd on the Rock" in 'strict' mode via the piano **(left)** and flute **(right)** were scanned within the frequency span from 4 Hz to 37 Hz utilizing the Duffing oscillator. Scale index (SI) values bottoming out downwards ($SI < 10^{-2}$) show weak periodic (or quasi-periodic) signal frequencies detected in EEG data of **Fp1, Fp2, F3, F7, F8** and **C3** channels.

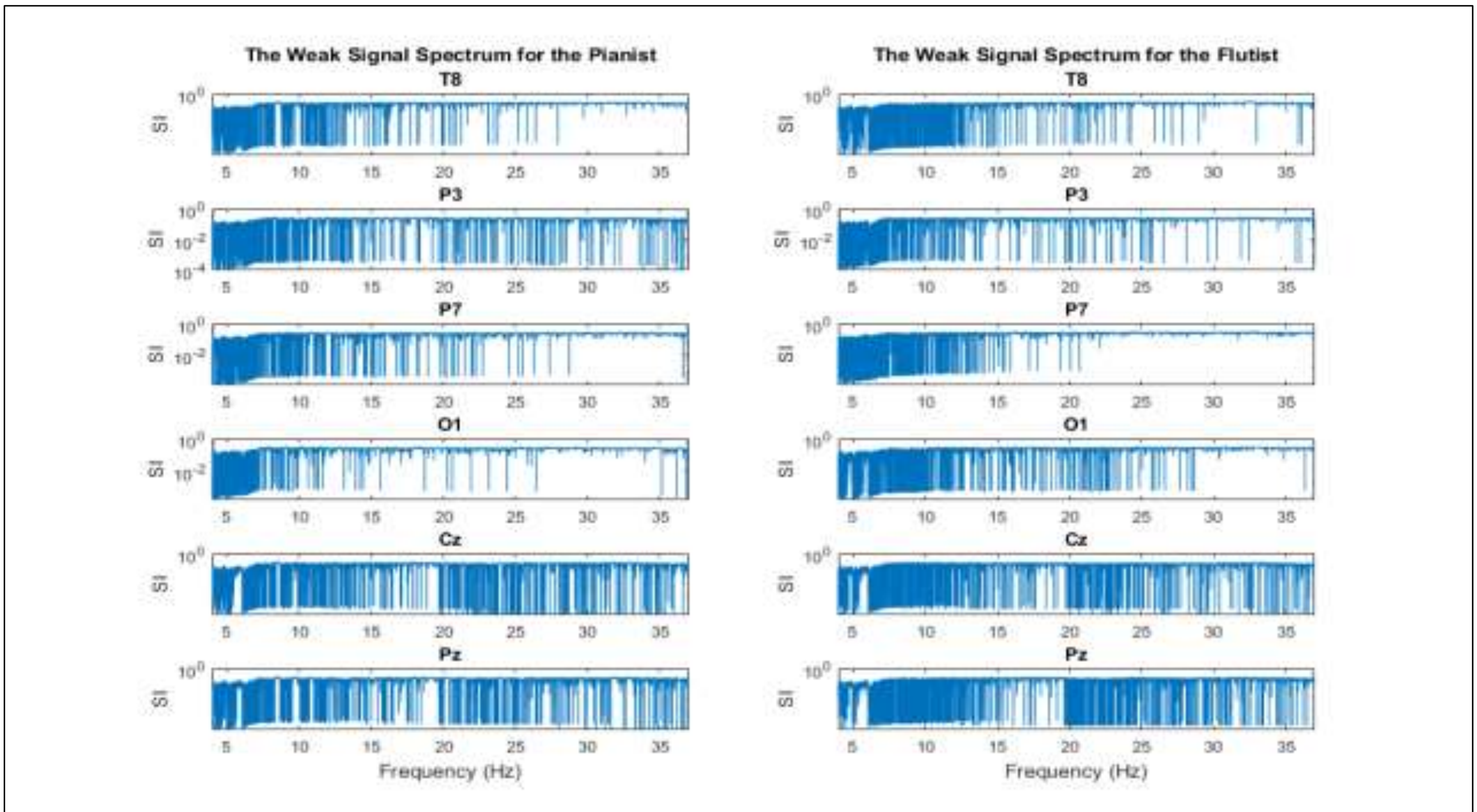

**Figure 5**: Spectrum of wavelet scale index parameters plotted against Duffing oscillator frequency: The EEG signals of two musicians performing Franz Schubert's "Shepherd on the Rock" in 'strict' mode via the piano **(left)** and flute **(right)** were scanned within the frequency span from 4 Hz to 37 Hz utilizing the Duffing oscillator. Scale index (SI) values bottoming out downwards ( $SI < 10^{-2}$) show weak periodic (or quasi-periodic) signal frequencies detected in EEG data of **T8, P3, P7, O1, Cz** and **Pz** channels.

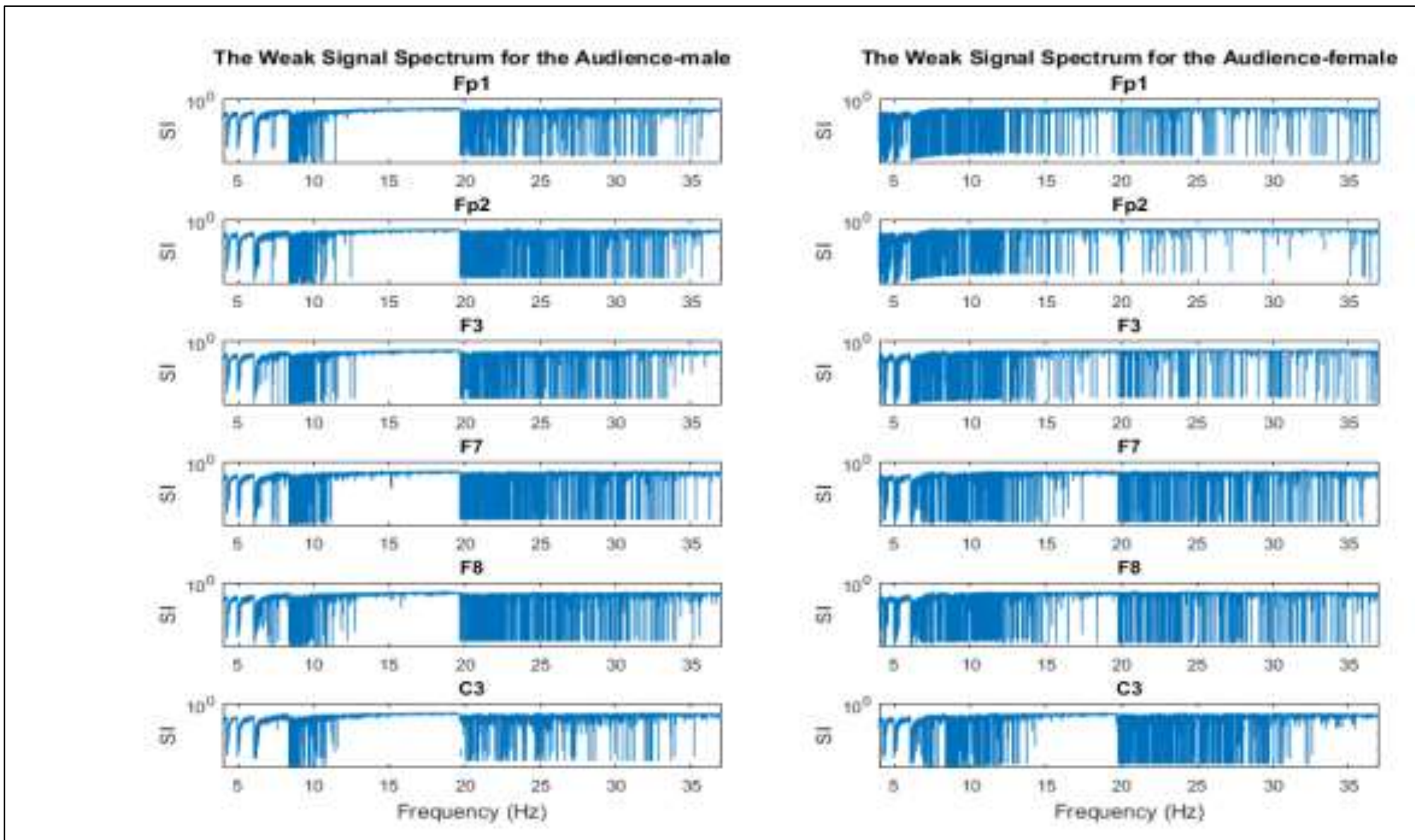

**Figure 6**: Spectrum of wavelet scale index parameters plotted against Duffing oscillator frequency: The EEG signals of the male audience (**left**) and female audience (**right**) were scanned within the frequency span from 4 Hz to 37 Hz utilizing the Duffing oscillator. Scale index (SI) values bottoming out downwards ($SI < 10^{-2}$) show weak periodic (or quasi-periodic) signal frequencies detected in EEG data of **Fp1, Fp2, F3, F7, F8** and **C3** channels.

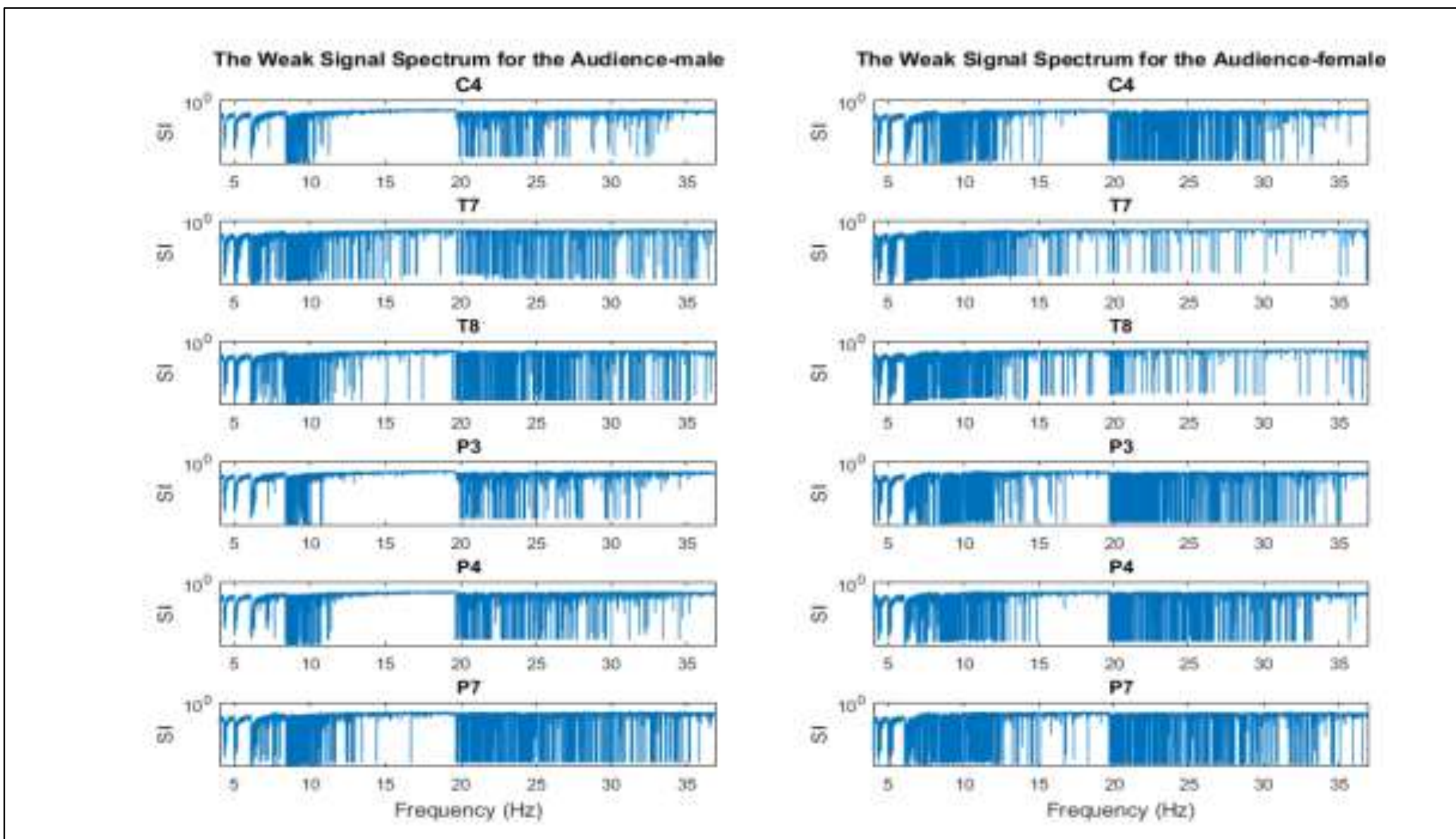

**Figure 7**: Spectrum of wavelet scale index parameters plotted against Duffing oscillator frequency: The EEG signals of the male audience (**left**) and female audience (**right**) were scanned within the frequency span from 4 Hz to 37 Hz utilizing the Duffing oscillator. Scale index (SI) values bottoming out downwards ( $SI < 10^{-2}$) show weak periodic (or quasi-periodic) signal frequencies detected in EEG data of **C4, T7, T8, P3, P4** and **P7** channels.

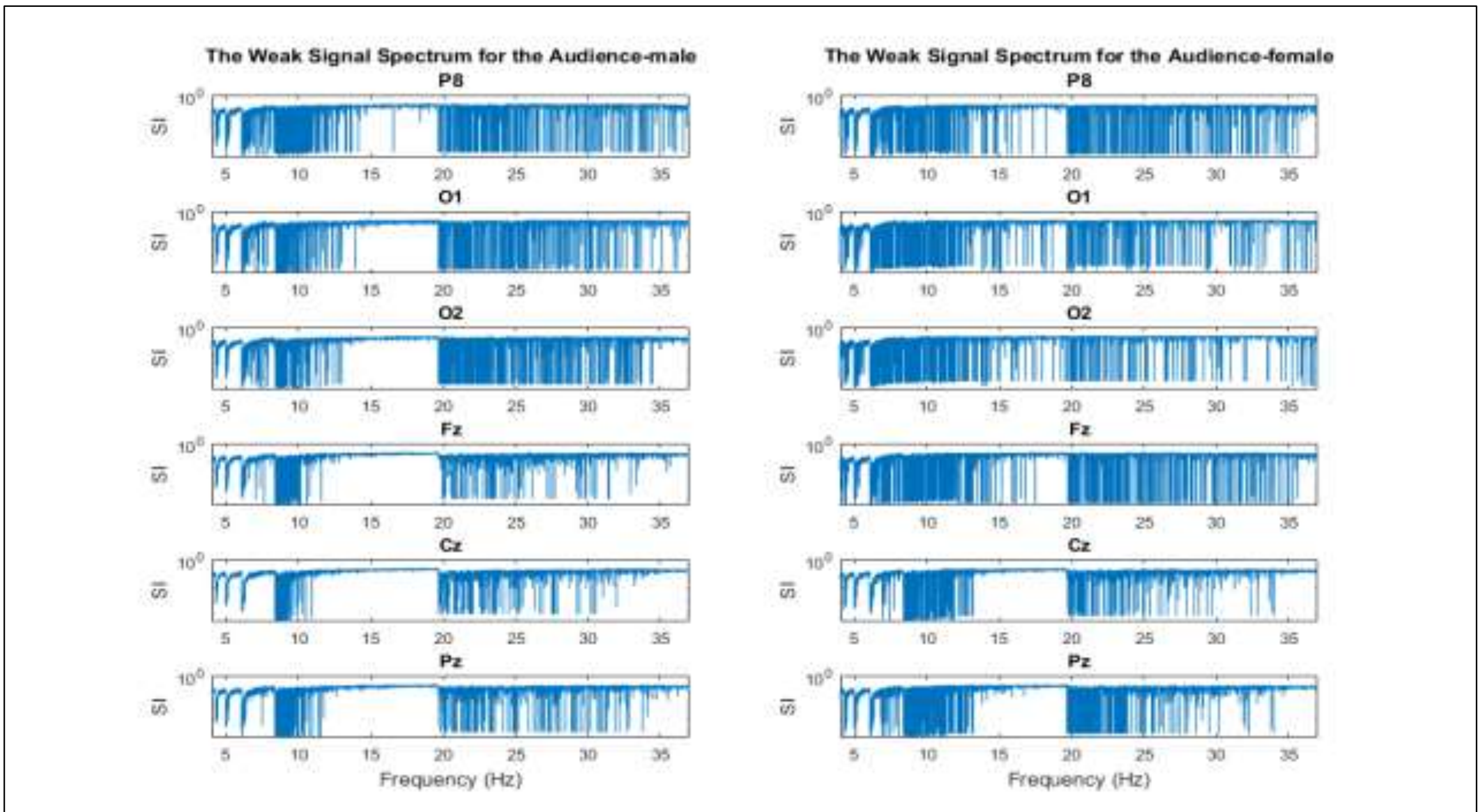

**Figure 8**: Spectrum of wavelet scale index parameters plotted against Duffing oscillator frequency: The EEG signals of the male audience (**left**) and female audience (**right**) were scanned within the frequency span from 4 Hz to 37 Hz utilizing the Duffing oscillator. Scale index (SI) values bottoming out downwards ($SI < 10^{-2}$) show weak periodic (or quasi-periodic) signal frequencies detected in EEG data of **P8, O1, O2, Fz, Cz** and **Pz** channels.

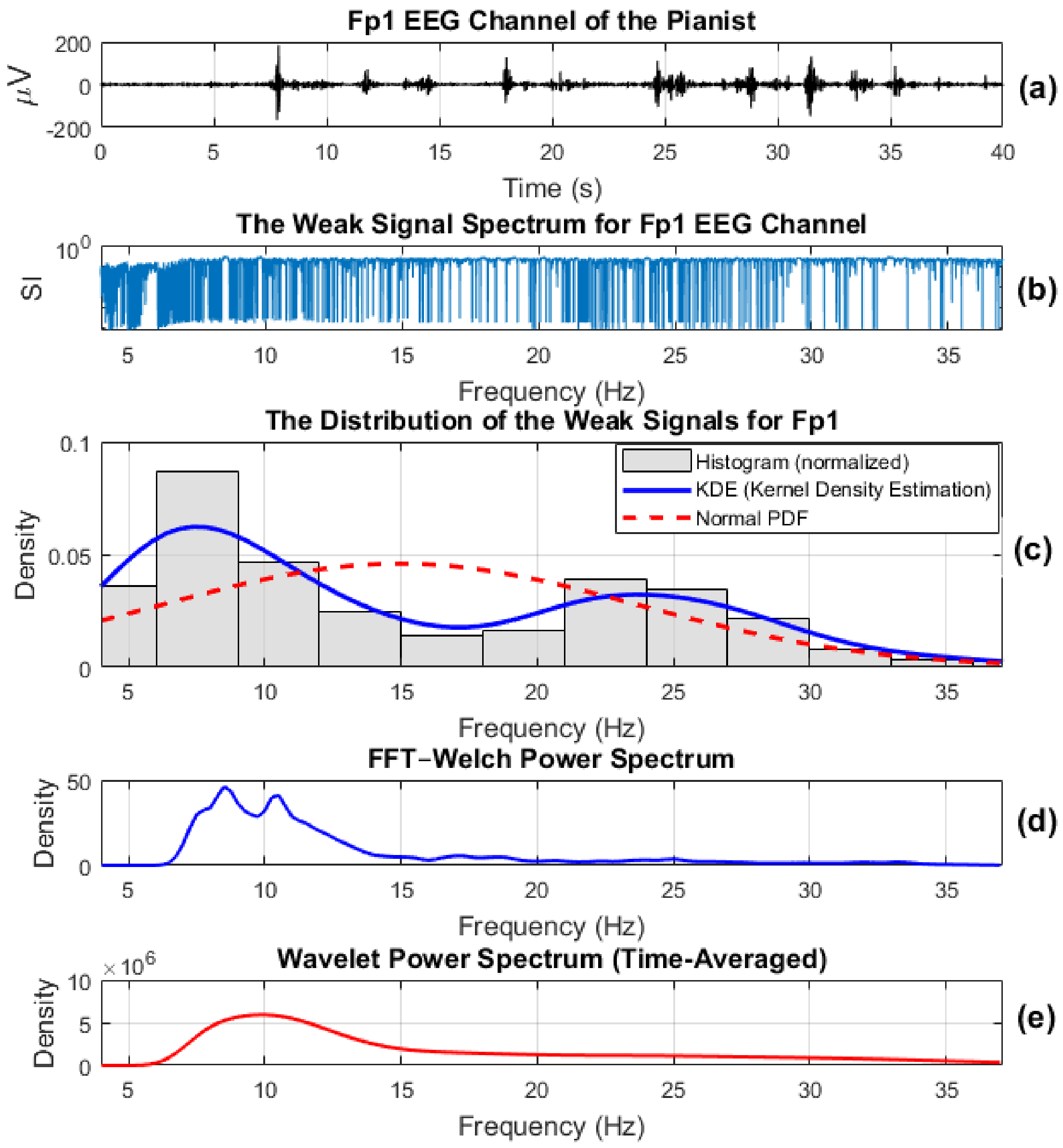

**Figure 9**: (**a**) EEG signal recorded from the **pianist**'s Fp1 channel during the first 40-second interval. [17]. (**b**) Weak signal spectrum in the 4–37 Hz frequency range for the Fp1 EEG channel. (**c**) Frequency distribution of weak signals, along with the corresponding histogram, normal distribution, and kernel density estimation. (**d**) Power spectrum of the Fp1 EEG signal obtained using the FFT–Welch method. (**e**) Wavelet power spectrum of the Fp1 EEG signal obtained by time-averaging the wavelet scalogram.

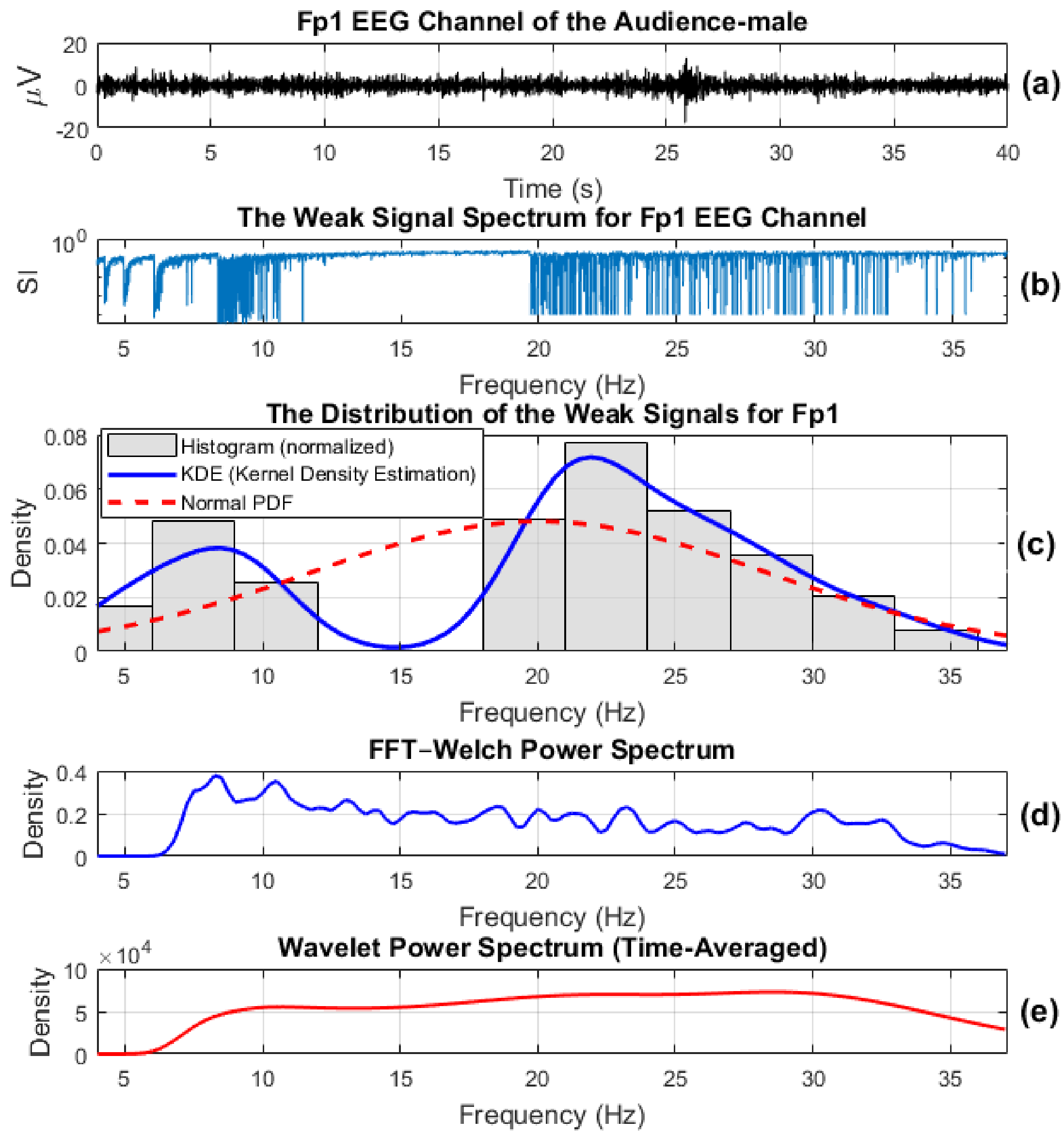

**Figure 10**: (**a**) EEG signal recorded from the **Audience-male**'s Fp1 channel during the first 40-second interval. [17]. (**b**) Weak signal spectrum in the 4–37 Hz frequency range for the Fp1 EEG channel. (**c**) Frequency distribution of weak signals, along with the corresponding histogram, normal distribution, and kernel density estimation. (**d**) Power spectrum of the Fp1 EEG signal obtained using the FFT–Welch method. (**e**) Wavelet power spectrum of the Fp1 EEG signal obtained by time-averaging the wavelet scalogram.

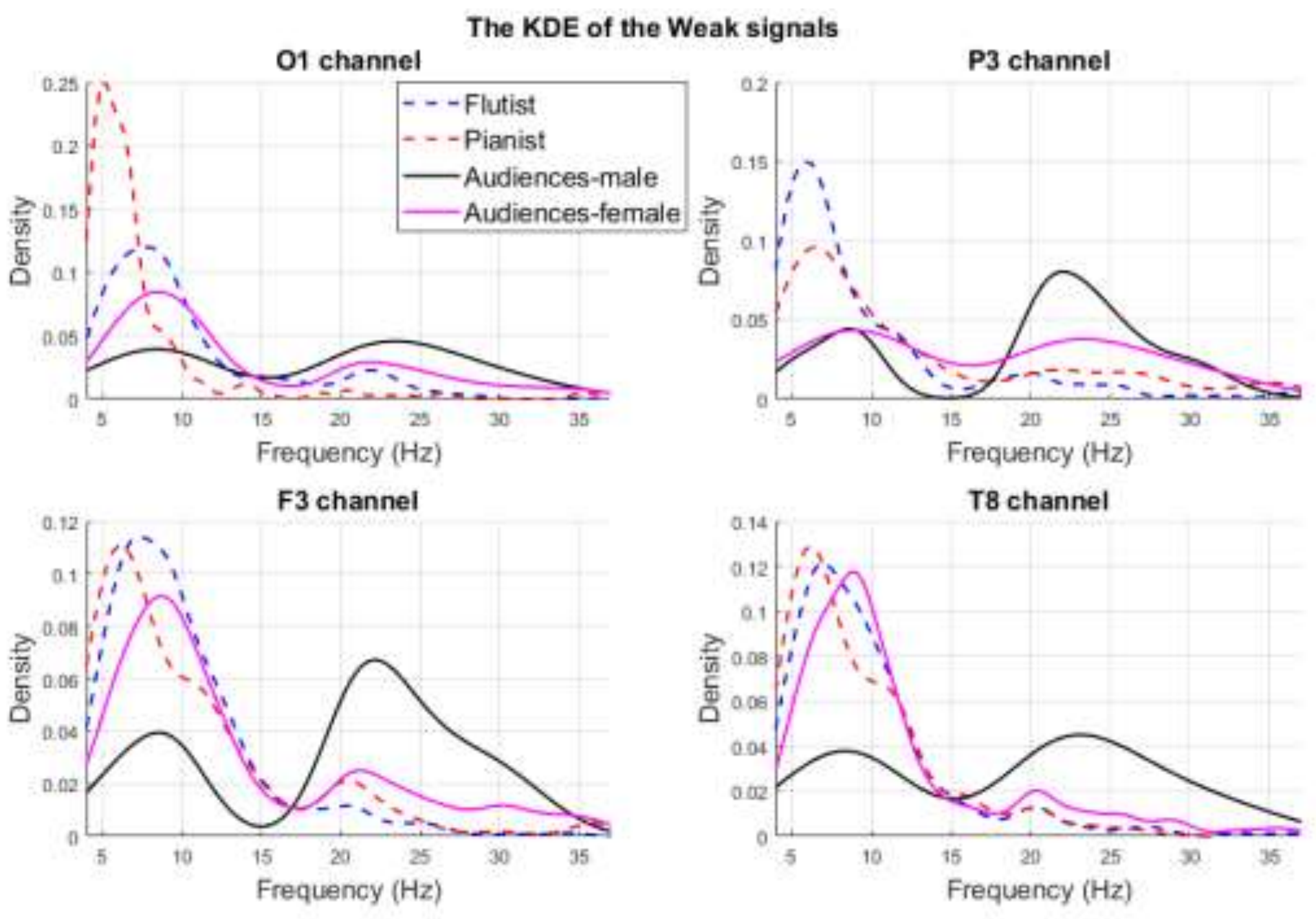

**Figure 11**: Comparison of the Kernel Density Estimation (KDE) for weak signals detected in EEG channels O1, P3, F3, and T8 for musicians and audiences.

### 3.2. The Brain Topological Map of the Weak Signals

The weak signals were visualised on the surface of the two-dimensional brain according to the above results. The 4–37 Hz frequency range was divided into four conventional EEG bands—Theta (4–8 Hz), Alpha (8–12 Hz), Low Beta (12–20 Hz), and High Beta (20–37 Hz)—and four corresponding brain topographic maps were generated. Colouring was done according to the number of weak signals detected. The regions shown in red indicate the high number of weak signals and the regions shown in blue indicate the low number of weak signals. **Figures 12, 13, 14,** and **15** present the topographic maps of detected weak signals in the EEG recordings of two musicians and two listeners, corresponding to the Theta (4–8 Hz), Alpha (8–12 Hz), Low Beta (12–20 Hz), and High Beta (20–37 Hz) frequency bands, respectively.

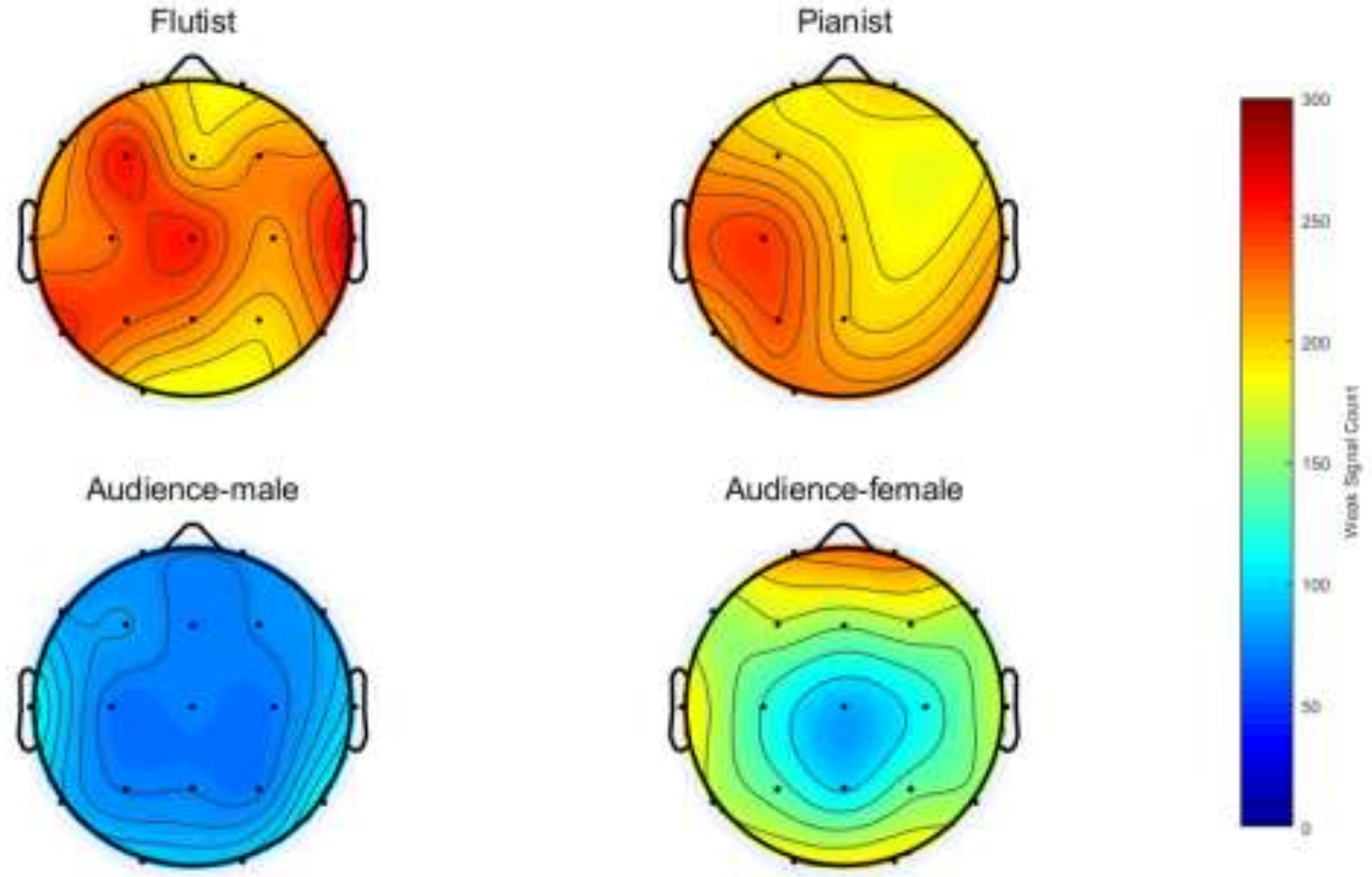

**The topological map of the weak signals between 4-8 Hz (Theta)**

**Figure 12**: The topological map of the weak signals with periodic or quasi-periodic structure that were detected in the EEG signals of two musicians and two audience members between <u>4 Hz and 8 Hz (Theta)</u> The black dots show the positions of the electrons according to the 10/20 system.

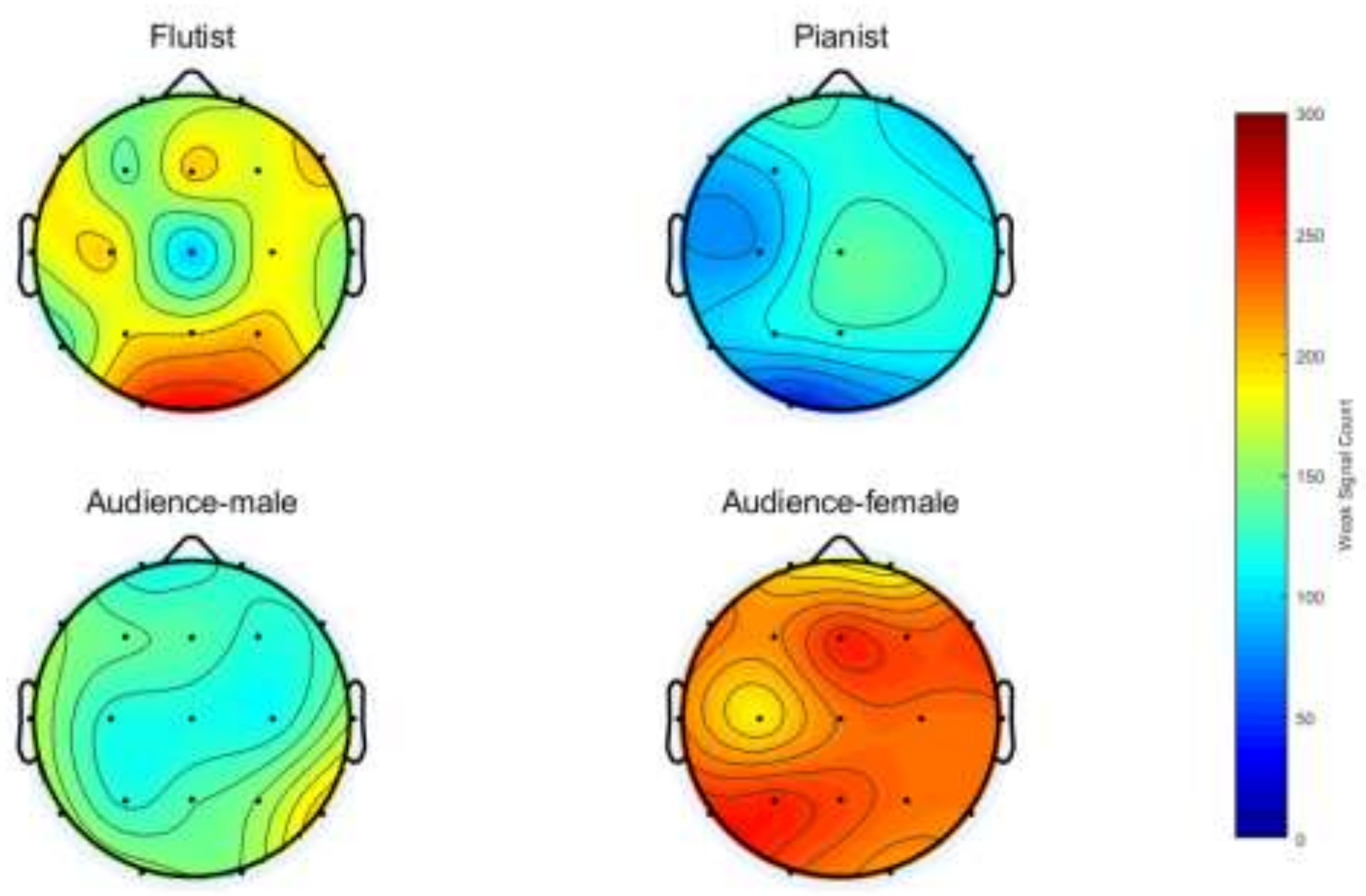

**The topological map of the weak signals between 8-12 Hz (Alpha)**

**Figure 13**: The topological map of the weak signals with periodic or quasi-periodic structure that were detected in the EEG signals of two musicians and two audience members between <u>8 Hz and 12 Hz (Alpha).</u> The black dots show the positions of the electrons according to the 10/20 system.

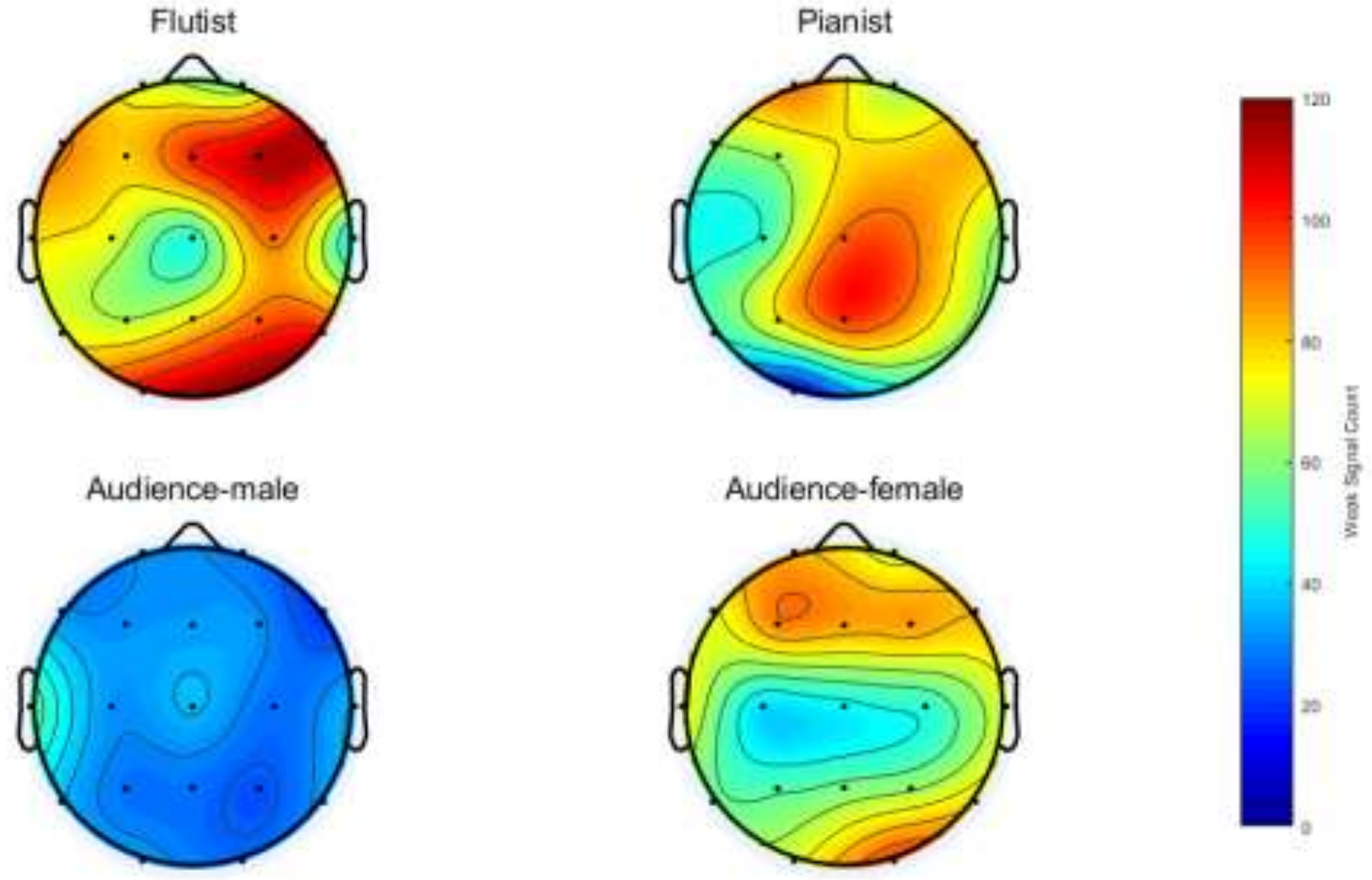

**The topological map of the weak signals between 12-20 Hz (Low Beta)**

**Figure 14**: The topological map of the weak signals with periodic or quasi-periodic structure that were detected in the EEG signals of two musicians and two audience members between <u>12 Hz and 20 Hz (Low Beta).</u> The black dots show the positions of the electrons according to the 10/20 system.

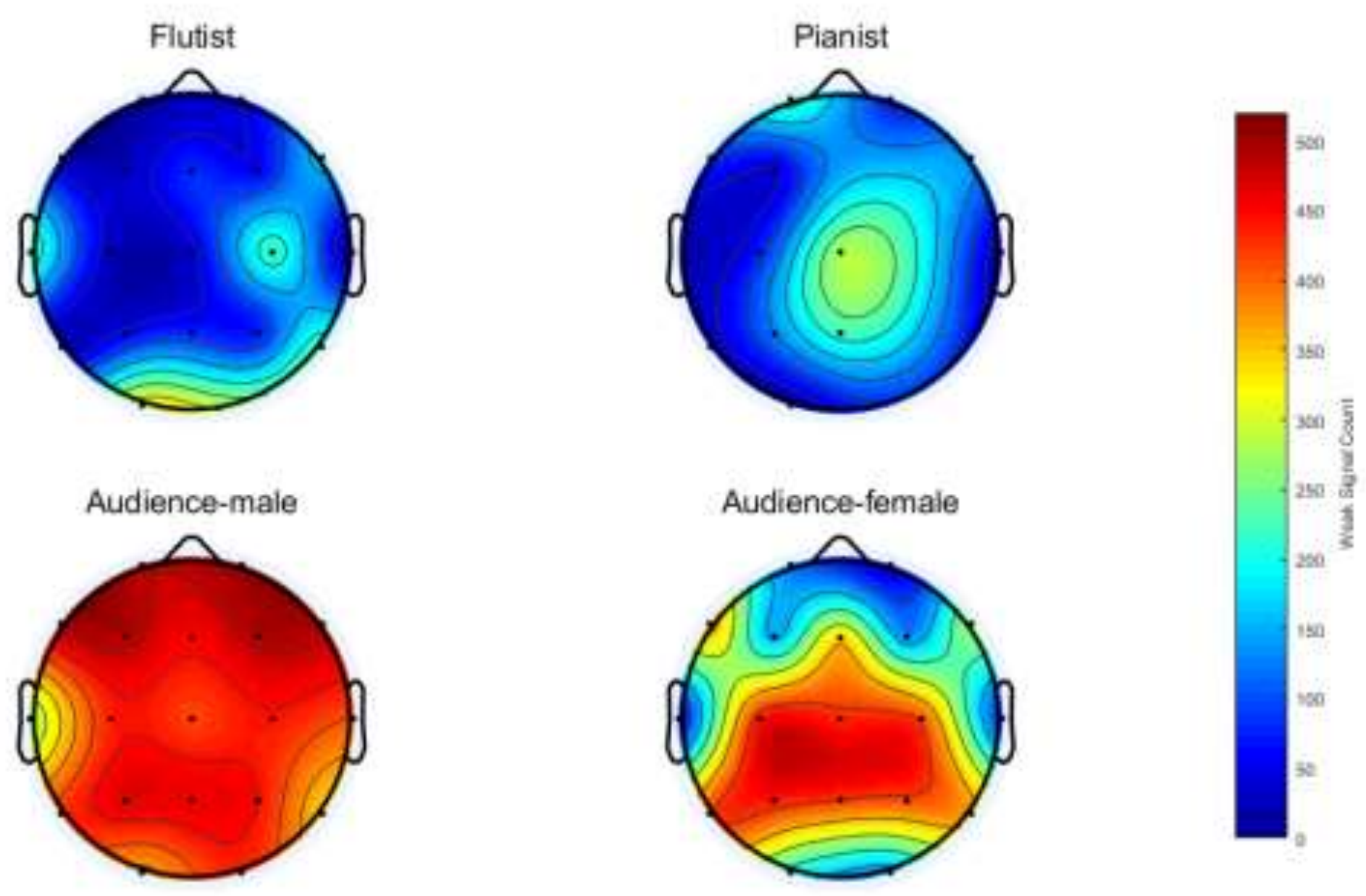

**The topological map of the weak signals between 20-37 Hz (High Beta)**

**Figure 15**: The topological map of the weak signals with periodic or quasi-periodic structure that were detected in the EEG signals of two musicians and two audience members between <u>20 Hz and 37 Hz (High Beta).</u> The black dots show the positions of the electrons according to the 10/20 system.

## 4. Discussion

The Duffing chaotic oscillator is an effective method in finding weak signals within EEG signals characterised by strong background noise [14-16]. Building on this capability, this study proposes a brain mapping approach based on the frequency distribution of postsynaptic potentials extracted from EEG signals using the Duffing oscillator.

EEG signals originate from synaptic potentials of pyramidal neurons in the cerebral cortex of the brain and are recorded by placing electrodes on the scalp. The post-synaptic potentials exhibit prolonged and consistent propagation times, sharing the same waveform, and are susceptible to both temporal and spatial summation [25-28]. Therefore, the Duffing oscillator system can detect post-synaptic potentials resulting from synchronous firing of pyramidal neurons in a region of the brain as periodic or quasi-periodic weak signals. Essentially, we assume that the weak signals detected in EEG signals represent the post-synaptic potentials of activated pyramidal neurons. Therefore, the amount of these weak signals serves as an indicator of the brain's activity level.

In this study, the Duffing oscillator was used to perform EEG brain mapping of two musicians (pianist and flutist) and two audience members (male and female) across the 4–37 Hz frequency range.

As can be seen from the **Figures 4-8,** there are significant differences in the frequency spectrum of weak signals found within the EEG signals of musicians and audience members. There are many more the weak signals in the EEG of the musicians than in the EEG of the audience members within theta band (4-8 Hz). **Figures 6**, **7** and **8** show that, unlike the musicians, the number of weak signals in the audience EEGs increases dramatically at frequencies above 19.5 Hz within the beta band (12–37 Hz).

**Figures 4 and 5** show that the weak signal spectra of the flutist and the pianist are generally similar, although there are small differences. However, their Fp1 and O1 channels show significant differences compared to each other. **Figures 6, 7 and 8** show that the weak signal spectra of the male audience and female audience are generally similar. However, between 12 Hz and 19.5 Hz, the weak signals' number within the EEG of the female audience is slightly higher than that of the male audience.

**Figures 12-15** illustrate the topographic maps based on the frequency distribution of detected weak signals in the EEG recordings of two musicians and two listeners, corresponding to the Theta (4–8 Hz), Alpha (8–12 Hz), Low Beta (12–20 Hz), and High Beta (20–37 Hz) frequency bands, respectively. In these brain maps, the musicians exhibit higher brain activity in the Theta and Low Beta bands, whereas the listeners show increased activity in the High Beta band. In contrast, no significant difference in activity levels is observed between musicians and listeners in the Alpha band. The following conclusions can be drawn from these brain maps: The musicians' brain activity levels are higher than the audience members within theta band (4-8 Hz), as shown in **Figure 12**. Therefore, it can be assumed that this is the frequency range which is in connection with the music playing activity. The brain activity level of audience members is quite high within high beta band (20-37 Hz), as shown in **Figure 15**. Therefore, it can be assumed that this frequency range is related to the enjoyment of listening to music.

**Figure 11** shows a comparison of the KDE of weak signals detected in EEG channels O1, P3, F3 and T8. The KDE of weak signals can be used to compare EEG channels. Also, it can be assessed whether there are synchronization states between certain regions of the brain by evaluating the peaks in the frequency intensities of these weak signals in KDE graphs.

The results show that the Duffing oscillator can be used as a scanner to obtain EEG brain maps. In addition, the Duffing oscillator has advantages according to other qEEG techniques [29-36] based on the 'Wavelet transform' or 'Fourier transform' due to its noise immunity. The Fourier and Wavelet transform is a technique based on the calculation of amplitudes. Therefore, as shown in **Figures 9(d)-(e)** and **10(d)-(e)**, these techniques are quite successful in showing levels of brain activity because EEG amplitude values are large at low frequencies. However, since EEG amplitudes are small at high frequencies, these techniques are not successful in showing the level of brain activity due to noise. Duffing oscillator-based brain mapping is quite successful in showing levels of brain activity at both low and high frequencies, as shown in **Figures 9(c)** and **10(c)**. Consequently, the findings show that Duffing oscillator–based weak signal detection is a powerful tool for brain mapping. When compared to Fourier- and wavelet-based techniques, the proposed method reveals brain activation maps with higher spatial detail. Thus, this technique can be used in neuroscience to study various pathological conditions and behavioural functions of the brain.

**Acknowledgment:** We would like to express our sincere gratitude to Professor Dr. **Henrik Jeldtoft Jensen**, Head of the Centre for Complexity Science, Department of Mathematics, Imperial College London, for his invaluable support and for granting us access to the EEG dataset originally collected by his group in collaboration with the Guildhall School of Music and Drama. We also thank Hardik Rajpal, Madalina Sas and Alberto Liardi, members of the Centre for Complexity Science, for their valuable discussions.

**Conflict of interest:** The authors have no Conflict of interest to disclose.

**Ethics Statement:** Ethical approval for the use of EEG signals in a previously published study [17] was given by the Guildhall School of Music and Drama Ethics Committee, Imperial College London, https://doi.org/10.3389/fpsyg.2018.01341 . Therefore, ethics committee approval was not required for this study.

**Data Availability Statement:** The data are available from the corresponding author upon reasonable request.